\documentclass[sigconf]{acmart}
\usepackage{longtable}
\usepackage{booktabs}
\usepackage{amsmath}
\usepackage{mathtools}
\usepackage{algorithm, tabularx}
\usepackage{algpseudocode}
\usepackage{amssymb}
\usepackage{multirow}
\usepackage{array}
\usepackage{makecell}
\usepackage{float}
\usepackage{chngcntr}
\usepackage{hhline}

\copyrightyear{2019} 
\acmYear{2019} 
\setcopyright{acmcopyright}
\acmConference[KDD '19]{The 25th ACM SIGKDD Conference on Knowledge Discovery and Data Mining}{August 4--8, 2019}{Anchorage, AK, USA}
\acmBooktitle{The 25th ACM SIGKDD Conference on Knowledge Discovery and Data Mining (KDD '19), August 4--8, 2019, Anchorage, AK, USA}
\acmPrice{15.00}
\acmDOI{10.1145/3292500.3330859}
\acmISBN{978-1-4503-6201-6/19/08}

\begin{document}
\title[MeLU: Meta-Learned User Preference Estimator for Cold-Start Recommendation]{
MeLU: Meta-Learned User Preference Estimator \\ for Cold-Start Recommendation
}

\author{Hoyeop Lee, Jinbae Im, Seongwon Jang, Hyunsouk Cho, Sehee Chung}
\affiliation{%
  \institution{Knowledge AI Lab., NCSOFT Co., South Korea}
}
\email{{hoyeoplee, jinbae, swjang90, dakgalbi, seheechung}@ncsoft.com}

\newcommand{\ours}{\textsf{MeLU}}
\newcommand{\eg}{{\it e.g.}}
\newcommand{\ie}{{\it i.e.}}
\newcommand{\etal}{{\it et al.}}
\newcommand{\etc}{{\it etc}}

\definecolor{ao_eng}{rgb}{0.0, 0.5, 0.0}
\newcommand{\comim}{\textcolor{ao_eng}}

\newcommand*\rot{\rotatebox{90}}

\makeatletter
\newcommand{\multiline}[1]{%
  \begin{tabularx}{\dimexpr\linewidth-\ALG@thistlm}[t]{@{}X@{}}
    #1
  \end{tabularx}
}
\makeatother

\begin{abstract}

This paper proposes a recommender system to alleviate the cold-start problem that can estimate user preferences based on only a small number of items.
To identify a user's preference in the cold state, existing recommender systems, such as Netflix, initially provide items to a user; we call those items evidence candidates. Recommendations are then made based on the items selected by the user.
Previous recommendation studies have two limitations: (1) the users who consumed a few items have poor recommendations and (2) inadequate evidence candidates are used to identify user preferences.
We propose a meta-learning-based recommender system called \ours\ to overcome these two limitations. 
From meta-learning, which can rapidly adopt new task with a few examples, \ours\ can estimate new user's preferences with a few consumed items.
In addition, we provide an evidence candidate selection strategy that determines distinguishing items for customized preference estimation. We validate \ours\ with two benchmark datasets, and the proposed model reduces at least 5.92\% mean absolute error than two comparative models on the datasets. We also conduct a user study experiment to verify the evidence selection strategy.

\end{abstract}


\maketitle

\section{Introduction}
\label{sec:intro}

Recommender systems can be generally classified as collaborative filter-based, content-based, or hybrid systems. Collaborative filter-based systems estimate user responses by collecting preference information 
from numerous users~\cite{linden2003amazon,sedhain2015autorec,he2017neural}. 
The predictions are built upon the existing ratings of other users who have similar ratings as the target user. However, such systems cannot handle new users (user cold-start) and new items (item cold-start) because of the lack of user-item interactions. Content-based systems are introduced~\cite{mooney2000content,narducci2016concept} to solve the cold-start problem.
Such systems use user profile information (\eg, gender, nationality, religion, and political stance) and the contents of the items to make recommendations. The systems might have a limitation suggesting the same items to the users who have similar contents regardless of items that user already rated.
Hybrid systems, which are based on a collaborative filter and utilize content information, are widely used in various applications~\cite{cheng2016wide,strub2016hybrid,kouki2015hyper}. 
However, these systems are unfit for a recommendation when the user-item interaction data are sparse. Moreover, due to privacy issues, collecting personal information is challenging, which might result in the user cold-start problem.

To avoid the user cold-start problem due to privacy issues, many web-based systems, such as Netflix, recommend items based on only minimal user information. Netflix initially presents popular movies and television programs to new users: we call these videos the \textit{evidence candidates}. Then, the user chooses the videos that he/she likes among the candidates. Afterward, the system recommends some programs based on the videos selected by the user. 
Recently, to improve performance, the recommendations have been made using deep learning methods~\cite{van2013deep, wei2017collaborative, li2017collaborative}; however, the cold-start problem remains for new users who rate only a few items.

Previous recommender systems are limited by two important issues.
First, a system should be able to recommend items to new users who have made a few ratings. New users leave the system when they receive poor recommendations initially~\cite{mcnee2006being}. However, the existing systems are not designed for users who rate only a few items. The previous system leverages the user profile information to improve the poor performance, but it does not solve the limitation. Consider two unemployed male users of a movie service who are in their twenties. One man watches a few science fiction movies, while the other man views a few horror movies. When gender, age, and occupation are provided as the user information, the recommender system might present a very similar movie lists to both men because a few movies cannot figure out their preferences.
Second, the existing systems do not provide reliable evidence candidates to estimate user preferences: they show popular items as evidence candidates. Spending time selecting evidence candidates might not be necessary because recommendations naturally become robust as user-item interaction increases. 
However, we have to deliberately select proper evidence candidates to improve the initial recommendations for new users.

This study proposes a meta-learning-based recommender system called \ours\ to resolve the preceding issues. Meta-learning focuses on improving classification or regression performance by learning with only a small amount of training data. 
A recommender system has similar characteristics with meta-learning, because it focuses on predicting a user's preferences based on only a small number of consumed items.
We consider the Model-Agnostic Meta-Learning (MAML) algorithm~\cite{finn2017model}, which allows estimation of customized preference directly based on an individual user's item-consumption history, even when only a small number of items have been consumed. In contrast to collaborative filter-based systems, which find other users who have similar ratings as the target user, the proposed system considers the items consumed by the target user only. 
Moreover, we suggest evidence candidate selection strategy for the MAML-based recommender system which can substantially enhance the initial recommendation performance for new users by selecting distinguishing items for customized preference estimation. 

Our study provides four main contributions. First, we alleviate the user cold-start problem by adopting the MAML concept in the recommender system.
Second, to the best of our knowledge, this study is the first to identify reliable evidence candidates to improve the initial recommendation performance of new users.
Third, \ours\ can provide personalized model to each user using their unique item-consumption history. Lastly, the proposed system and evidence candidate selection strategy are validated with two benchmark datasets and a user survey.

The remainder of this paper is organized as follows. We briefly describe meta-learning and the relationship between meta-learning and the proposed recommender model in Section~\ref{sec:prel}. We present the proposed system and evidence candidate selection strategy in Section~\ref{sec:model}. Section~\ref{sec:exp} evaluates the recommendation performance of the proposed method by means of two comparative methods for benchmark datasets. In addition, through user study, we compare the proposed model-based evidence candidates with popularity-based evidence candidates in Section~\ref{sec:exp2}. Finally, we conclude this paper and discuss future research in Section~\ref{sec:con}.

\section{Meta-Learning}
\label{sec:prel}

Meta-learning, also called learning-to-learn, aims to train a model that can rapidly adapt to a new task which is not used during the training with a few examples~\cite{vilalta2002perspective}. Meta-learning is inspired by the human learning process, which can quickly learn new tasks based on a small number of examples. Meta-learning can be classified into three types: metric-based, memory-based, and optimization-based meta-learning. Previous researches~\cite{koch2015siamese, santoro2016meta, vinyals2016matching, snell2017prototypical} on metric-based and memory-based meta-learning are concentrated on classification problems; the work in~\cite{vartak2017meta} took a concept of the metric-based meta-learning algorithm in the recommender system to predict whether a user consumes an item or not. The system could make a recommendation based on item-consumption but cannot provide a personalized model to users.
By contrast, we consider taking a concept of optimization-based meta-learning~\cite{finn2017model, li2017meta} to recommender system, which can serve a personalized recommender model by reflecting item-consumption of each user.

\begin{figure}
  \centering
  \includegraphics[width=50mm]{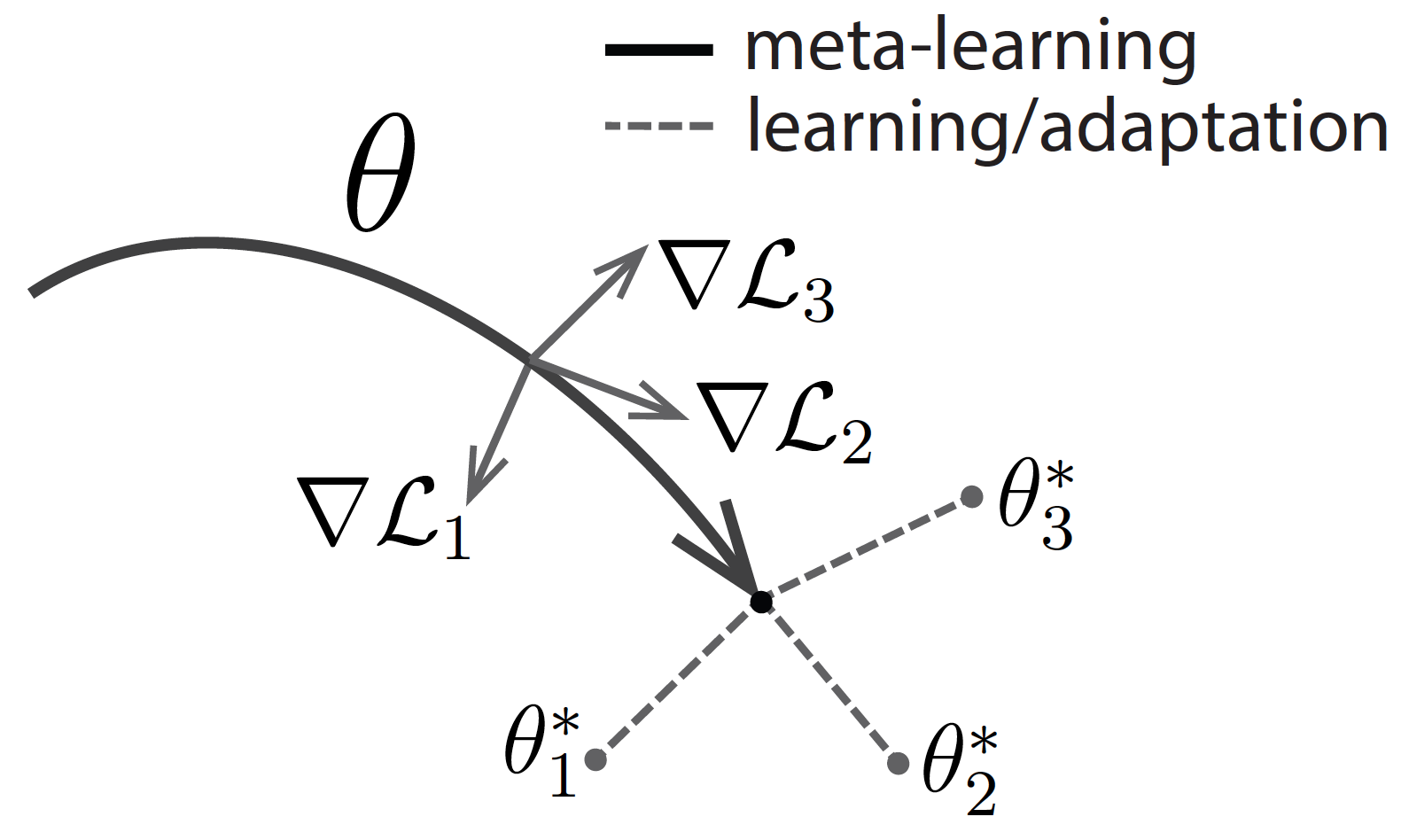}
  \caption{Diagram of the optimization-based meta-learning algorithm~\cite{finn2017model}.}
  \label{fig:maml}
\end{figure}

The optimization-based meta-learning algorithm considers the model $f$ and the distribution over task $p(\mathcal{T})$. The algorithm attempts to find desirable parameter $\theta$ of model $f$, as shown in Figure~\ref{fig:maml}. 
The optimization-based meta-learning algorithm performs local and global updates.
From the random initial parameter $\theta$ (the starting point of the black arrow), the algorithm samples several tasks from the distribution over task $\mathcal{T}_i\sim p(\mathcal{T})$. The algorithm \textit{locally} updates parameter $\theta$ to $\theta_{i}$ by gradient $\nabla_\theta \mathcal{L}_{i}(f_\theta)$ for each tasks $i=1,\cdots, T$, where $T$ is the number of sampled tasks and $\mathcal{L}_{i}(f_\theta)$ denotes the training loss on task $i$ with parameter $\theta$. The local updates are represented as gray arrows in the figure. After local updates, for all sampled tasks, the algorithm \textit{globally} updates parameter $\theta$ based on $\mathcal{L}'_{i}(f_{\theta_i})$, which is the test loss on task $i$ with parameter $\theta_i$, so that the globally updated parameter fits into the various tasks.

The optimization-based meta-learning algorithm uses two sets, namely, the support set and query set, for each task. The support set and query set are used for calculating the training loss and test loss on each task, respectively. During the local update, the algorithm adjusts the parameter of the model on each support set (learning process). During the global update, the algorithm trains the parameter to minimize the losses with the adapted parameters on the query sets (learning-to-learn process). When the learning-to-learn process reaches a termination condition for previous tasks, the algorithm only accepts a support set for a new task. Using the support set, the model can adapt to the new task. Note that, the algorithm allows parameters for each task not to be stored; instead, the parameters are calculated via support sets.

We regard each task as estimating a user's preferences in the recommender system. From this inspiration, we propose a MAML-based recommender system that can rapidly estimate a new user's preferences based on only a few user-item interactions.
The MAML-based recommender system accounts for the fact that different users have different optimal parameters. Therefore, our MAML-based recommender system provides each user personalized recommendations based on their unique item-consumption history. Consider the two male users from Section~\ref{sec:intro}. The item-consumption history of the first user includes science fiction movies, and that of the second user includes horror movies. Since our recommender system considers movies watched by individual users, the system will recommend science fiction movies or horror movies to each user.
Moreover, our model can provide items with large gradient as reliable evidence candidates. 
An item has a large gradient for item-consumption history if a rating of that item by a new user provides the system with substantial information about the preferences of the user. 

\section{Method}
\label{sec:model}

In this section, we describe the details of Meta-Learned User preference estimator (\ours).
We first introduce a user preference estimatior that consists of decision-making layers and output layer with user and item embeddings. Next, we design a personalized user preference estimation model based on MAML that can rapidly adapt to new tasks (new users). Finally, we suggest an evidence candidate selection strategy based on the personalized user preference estimation model.

\subsection{User Preference Estimator}

\begin{figure}
  \centering
  \includegraphics[width=80mm]{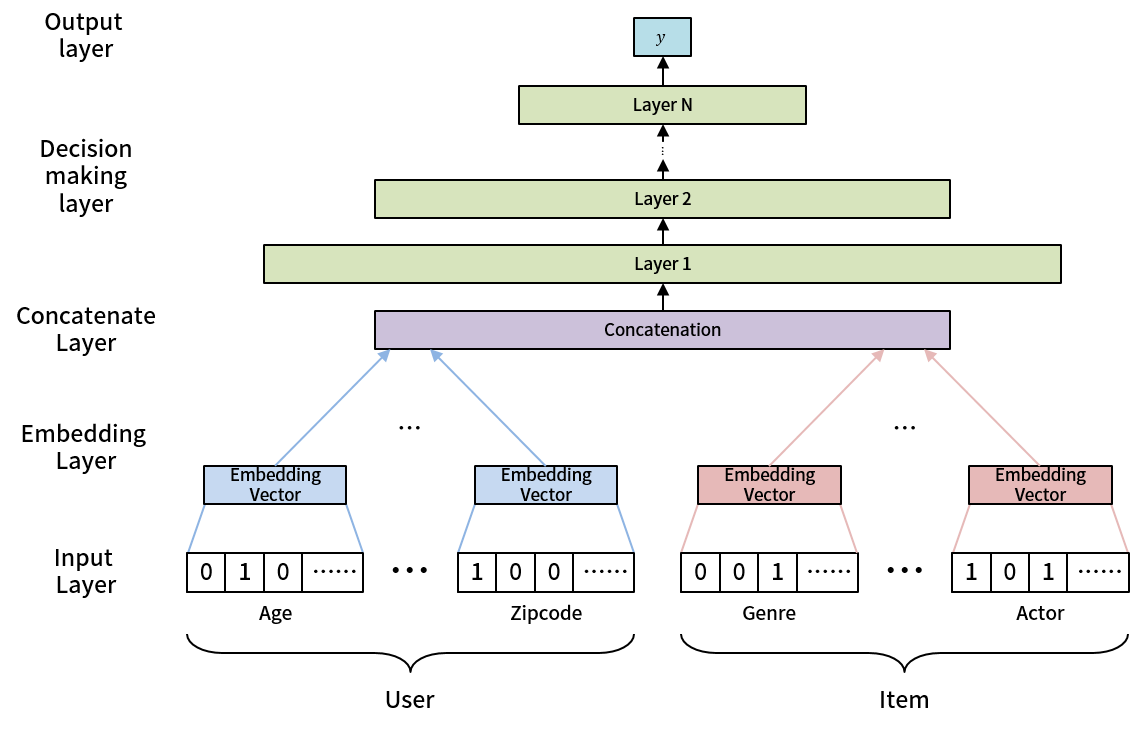}
  \caption{User preference estimatior.}
  \label{fig:framework}
\end{figure}

The proposed user preference estimation model for the recommender system is shown in Figure~\ref{fig:framework}. The model takes user content (\eg, age and occupation) and item content (\eg, genre and publication year) as the input. First, the proposed model performs embedding processes based on the input and concatenates the embedded vectors. Second, from the embedding, we model the decision-making process by means of a multilayered neural network, which is widely used in recent recommendation research~\cite{cheng2016wide, he2017neural}.

We employ the embedding process to extract useful features to estimate user preferences from the contents, because the previous research showed that the embedding process improves the performance in the recommender system~\cite{barkan2016item2vec, zhang2016collaborative}. Our model embeds user and item contents in a similar way as in previous work~\cite{cheng2016wide}.
For categorical content, we generate each content embedding and use the concatenated embeddings as shown in Figure~\ref{fig:framework}. When the number of user contents is $P$, we define the embedding vector $U_i$ for user $i$ as follows.
\begin{equation}
    U_i = \left[ e_{i1}c_{i1};\, \cdots; e_{iP}c_{iP} \right]^\intercal
    \label{eq:user_emb}
\end{equation}
where $c_{ip}$ is a $d_p$-dimensional one-hot vector for categorical content $p\in \{1,\, \cdots, P\}$ of user $i$ and $e_{ip}$ represents the $d_e$-by-$d_p$ embedding matrix for the corresponding categorical content of the user. $d_e$ and $d_p$ are the embedding dimension and the number of categories for content $p$, respectively.
The items are embedded in the same way. The embedding vector $I_j$ for item $j$ can be described as
\begin{equation}
    I_j= \left[ e_{j1}c_{j1};\, \cdots; e_{jQ}c_{jQ} \right]^\intercal,
    \label{eq:item_emb}
\end{equation}
where $c_{jq}$ is a $d_q$-dimensional one-hot vector for categorical content $q\in \{1,\, \cdots, Q\}$ of item $j$ and $e_{jq}$ represents the $d_e$-by-$d_q$ embedding matrix for the corresponding categorical content of the item. $Q$, $d_e$ and $d_q$ are the number of item contents, embedding dimension, and the number of categories for content $q$, respectively. If there exist continuous contents, the input nodes of the contents are connected directly to the concatenate layer without passing through the embedding layer.

User preference estimation is conducted by the decision-making and output layers of our model. The embedding dimensions for a user and an item may be different (\ie, $d_e\times p\neq d_e\times q$). Thus, we cannot use a generalized matrix factorization layer~\cite{he2017neural}, which requires equal embedding dimensions. Therefore, we construct the decision-making layer as a $N$-layer fully-connected neural network. The output layer, which is the subsequent layer of the decision-making layers, estimates user preferences that could be ratings, implicit feedback~\cite{kelly2003implicit}, or dwell times~\cite{yi2014beyond}. 
The layers can be formulated as
\begin{equation}
    \begin{split}
        x_0 & = [U_i; I_j], \\
        x_1 & = a(\mathbf{W}_1^\intercal x_0+b_1), \\
            & \vdots\\
        x_N & = a(\mathbf{W}_N^\intercal x_{N-1}+b_N),\\
        \hat{y}_{ij} & = \sigma (\mathbf{W}_{o}^\intercal x_N +b_{o}),
    \end{split}
    \label{eq:decisionlayer}
\end{equation}
where $\mathbf{W}_n$ and $b_n$ are the weight matrix and bias vector for the $n$-th decision-making layer, and $\mathbf{W}_o$ and $b_o$ are those for the output layer. $\hat{y}_{ij}$ is user $i$'s estimated preference for item $j$, and $a$ and $\sigma$ denote the activation functions of the decision-making layer and output layer, respectively. We use rectified linear unit (ReLU)~\cite{nair2010rectified} as the activation function $a$. The activation function $\sigma$ depends on how user preference is defined. If the goal is to estimate ratings and dwell times, a linear function might be appropriate, whereas the sigmoid function can be used in the case of implicit feedback.

\begin{figure*}
  \centering
  \includegraphics[width=0.9\textwidth]{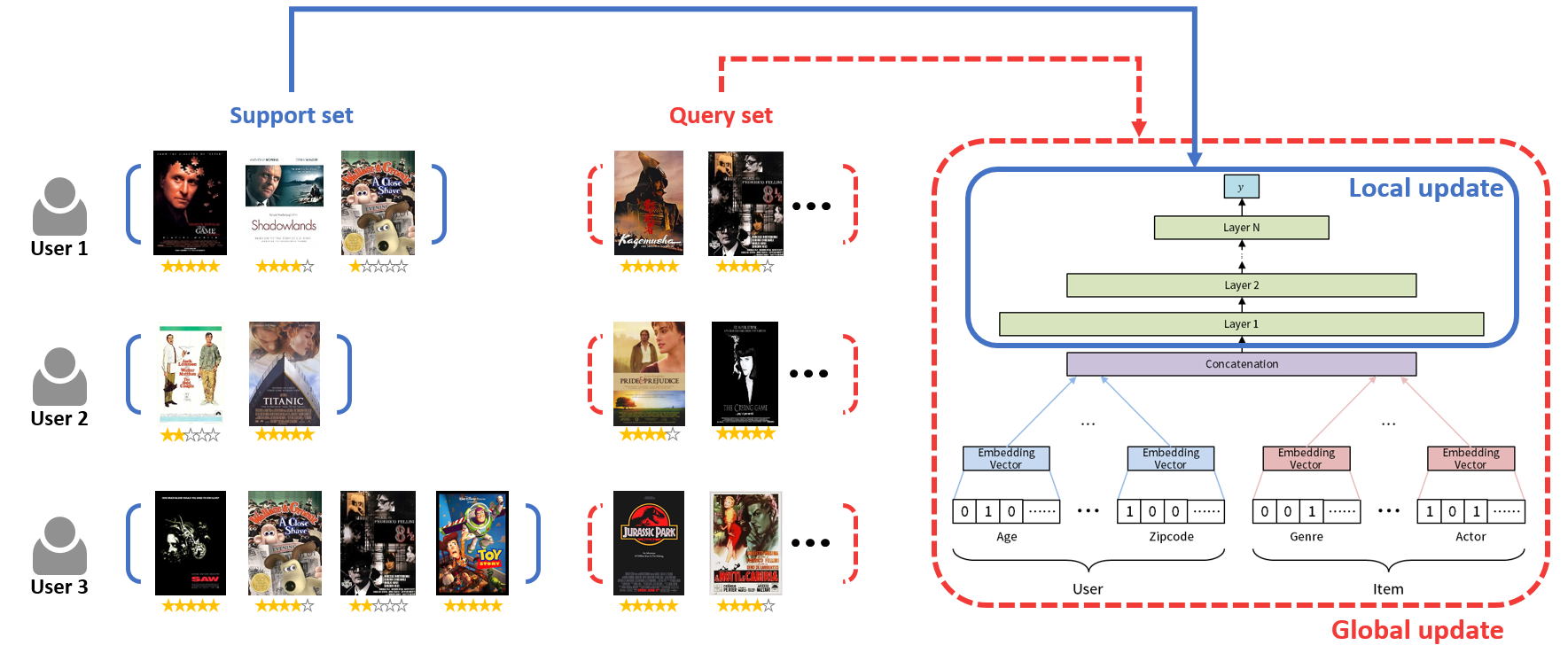}
  \caption{The proposed model, \ours, receives users' item histories and preferences. Then, the decision-making layers and the output layer (marked with a blue box) are updated based on the support set of each user; that is, the proposed model locally updates the user's decision-making process based on each user's item-consumption pattern. After local updates for the users, all layers (marked with a red-dashed box) are globally updated based on the query sets.}
  \label{fig:our_maml_framework}
\end{figure*}

\subsection{Meta-Learned User Preference Estimator}

Inspired by the concept of optimization-based meta-learning, we put this concept into \ours\ to reflect personalized user preferences with only a few user-item interactions.
Figure~\ref{fig:our_maml_framework} shows how the model employs users' item-consumption history, which is not directly considered in the previous model~\cite{cheng2016wide}. 
Our model considers a user's item-consumption history and preferences, and these inputs act as the support set for the local update. In other words, to reflect a user's tastes, the model updates the parameters in the decision-making layers and the output layer (\ie, $\mathbf{W}$ and $b$) based on the user's unique item-consumption history. We do not update the embedding vectors for the user and items during the local update to ensure the stability of the learning process, which means that our model assumes users and items do not change, only the users' thoughts change as they interact with the items.
Furthermore, unlike MAML, we do not limit the length of the item-consumption history. We extend the idea of the matching network, which is one of the most famous meta-learning algorithms and shows good performance even when the length of the support set (\ie, length of the item-consumption history) is not fixed~\cite{cai2018memory}.

Algorithm~\ref{algo:maml} shows the detailed personalization procedure of the user preference estimation model. First, the parameters in Eqs.~\ref{eq:user_emb}, \ref{eq:item_emb}, and \ref{eq:decisionlayer} (lines 1-2) are randomly initialized. As the model updates only the parameters in Eq. \ref{eq:decisionlayer} during the local update, we denote these parameters separately. 
Second, the model randomly samples the batch of users (line 4). Third, the model locally updates the parameters for user $i$ in the batch by backpropagating the following loss function (lines 6-8):
\begin{equation}
    \mathcal{L}_i=\frac{1}{|H_i|}\sum_{j\in H_i} (y_{ij} - \hat{y}_{ij})^2 \nonumber
\end{equation}
where $H_i=\{j|\text{ item } j \text{ consumed by user } i\}$ is a set of items consumed by user $i$, \ie, the support set, and $y_{ij}$ is user $i$'s actual preference for item $j$. This local update can be considered to be an iteration for personalization, which can be repeated several times. 
Finally, the algorithm globally updates $\theta_1$ and $\theta_2$ based on loss function $\mathcal{L}'_{i}$ for corresponding new item-consumption history $H'_i$ (\ie, the query set) with the personalized parameters (line 10). This process aims to find the desirable parameters that achieve good recommendation performance after a few local updates for all users. The algorithm repeats these processes until $\theta_1$ and $\theta_2$ converge. 
When recommending items to a user, \ours\ conducts local updates based on the user's item-consumption history and calculates the preferences for all items that the user has not rated. After that, the model recommends top favored items to the user.

\begin{algorithm}[t]
    \begin{algorithmic}[1]
        \caption{Model-Agnostic Meta-Learning for User Preference Estimator} \label{algo:maml}
        \Require $\alpha, \beta$: step size hyperparameters
        \State randomly initialize $\theta_1$ (parameters in Eqs.~\ref{eq:user_emb} and \ref{eq:item_emb})
        \State randomly initialize $\theta_2$ (parameters in Eq.~\ref{eq:decisionlayer}) 
        \While{not converge}
            \State sample batch of users $B\sim p(\mathcal{B})$
            \For{user $i$ in $B$}
                \State set $\theta_2^i = \theta_2$
                \State evaluate $\nabla_{\theta_2^i} \mathcal{L}_{i}(f_{\theta_1, \theta_2^i})$
                \State local update $\theta_2^i \gets \theta_2^i - \alpha \nabla_{\theta_2^i} \mathcal{L}'_{i}(f_{\theta_1, \theta_2^i})$
            \EndFor
            \State \multiline{%
                global update $\theta_1 \gets \theta_1 - \beta \sum_{i \in B} \nabla_{\theta_1} \mathcal{L}'_{i}(f_{\theta_1, \theta_2^i})$ \\
                $\mspace{104mu}\theta_2 \gets \theta_2 - \beta \sum_{i \in B} \nabla_{\theta_2} \mathcal{L}'_{i}(f_{\theta_1, \theta_2^i})$
            }
        \EndWhile
    \end{algorithmic}
\end{algorithm}

\subsection{Evidence Candidate Selection}

We suggest an evidence candidate selection strategy based on the \ours. The strategy identifies the distinguishing items that can be used to quickly analyze individual preferences of new users in the system.
In our model, the larger the average Frobenius norm of whole users' gradient for personalization is (\ie, the gradient of the local update, $\Vert\nabla_{\theta_2^i} \mathcal{L}_{i}(f_{\theta_1, \theta_2^i})\Vert_F$), the better the distinction among users' preferences. When calculating the gradient, we modify $\mathcal{L}_i$ as $\mathcal{L}_i/|\mathcal{L}_i|$, where $|\cdot|$ denotes the absolute value of the input to backpropagate the unit error. 
Although the items with large gradient are useful for identifying user preferences, a proper assessment might be difficult when the user does not know about the items. Therefore, we also consider the users' awareness of the items.
We assume that the more frequent user-item interaction is, the better aware users are of the item.
Therefore, for each item, we calculate the gradient-based value and popularity-based value as the average Frobenius norm and the number of interactions for each item using the whole existing user-item pairs, respectively.
To scale the units of two values, we normalize the values to range from zero to one and assign the score to each item by multiplying two normalized values.
Finally, we define the top $k$ items with the highest scores as evidence candidates. 
Note that, the scores can vary when entering new user, which directly effects the average Frobenius norm and user-item interactions.

\section{Experiments}
\label{sec:exp}

\subsection{Experimental Design}
We considered two datasets: MovieLens 1M~\cite{harper2016movielens} and Bookcrossing~\cite{ziegler2005improving}. 
Both datasets provide basic user and item information, such as user's age and publication year, and the datasets have explicit feedback information.
We divided the items and users into two groups (existing/new) to evaluate the performance under item-cold-start and user-cold-start condition. We prepared four partitions for each dataset: 1) existing items for existing users, 2) existing items for new users, 3) new items for existing users, and 4) new items for new users.
The following preprocessing was performed for each dataset:

\begin{itemize}
    \item \textbf{MovieLens 1M}: We collected additional movie contents from IMDb\footnote{\url{https://www.imdb.com/}}. 
    We divided the movies into movies released before 1997 and after 1998 (approximately 8:2). We regarded movies released before 1997 as existing items and movies released after 1998 as new items, and we randomly selected eighty percent of the users as existing users and the rests as new users.
    \item \textbf{Bookcrossing}: 
    We divided the books into those released before 1997 and those released after 1998 (approximately 5:5). We regarded books released before 1997 as existing items and books released after 1998 as new items, and we randomly selected one-half of the users as existing users and the remainder as new users. 
\end{itemize} 

 In each partition, we included only users whose item-consumption history length is between thirteen and one hundred. For each user, ten random items from the history were used as the query set ($H'_i$), and the remaining items were used as the support set ($H_i$) in Algorithm~\ref{algo:maml}, \ie, the length of the item-consumption history ranges from three to ninety. It was used to determine whether the model can learn consumers' preferences regardless of the length of the item-consumption history.
 The characteristics of the two datasets are summarized in Table~\ref{tab:data_sum}.

\begin{table}[t]
\centering
\caption{Basic statistics and used contents of the MovieLens and Bookcrossing dataset.}
\label{tab:data_sum}
\begin{tabular}{c|c|c}
    \hline
    \textbf{}           & \textbf{MovieLens} & \textbf{Bookcrossing} \\ 
    \hline \hline
    Number of users     & 6,040              & 278,858 \\ 
    Number of items     & 3,706              & 271,379 \\ 
    Number of ratings   & 1,000,209          & 1,149,780 \\ 
    Sparsity            & 95.5316\%          & 99.9985\% \\ 
    \hline 
    User contents       & \makecell{Gender, Age, \\ Occupation, Zip code}  & \makecell{Age} \\ 
    \hline 
    Item contents       & \makecell{Publication year, \\ Rate, Genre, \\ Director, Actor}  & \makecell{Publication year, \\ Author, Publisher} \\ 
    \hline 
    Range of ratings    & 1 $\sim$ 5         & 1 $\sim$ 10 \\ 
    \hline
\end{tabular}
\end{table}

The proposed model, \ours, was tested with the following structure. Two layers were used for the decision-making layers with 64 nodes each, and the dimension of all embedding vectors was set to 32. We set the step sizes $\alpha$ and $\beta$ to $5\times10^{-6}$ and $5\times10^{-5}$ for the MovieLens 1M dataset and to $5\times10^{-5}$ and $5\times10^{-4}$ for the Bookcrossing dataset. The number of local updates was varied from one to five. The batch size and the maximum number of epochs of Algorithm~\ref{algo:maml} were set to 32 and 30, respectively. All experiments were implemented with Pytorch~\cite{paszke2017automatic}. The source code of the proposed model is available online\footnote{\url{http://github.com/hoyeoplee/MeLU}}.

We compared the proposed method with two models: Pairwise Preference Regression (PPR)~\cite{park2009pairwise} and Wide \& Deep~\cite{cheng2016wide}. PPR estimates user preferences via bilinear regression with the one-hot user and item content vectors. Wide \& Deep predicts whether the user likes an item, but we modified it into a regression model that estimates preferences. The structure of the Wide \& Deep model was the same as that of our user preference estimation model. We did not consider the collaborative-filtering-based approaches for comparison because they cannot estimate preferences for new users and new items.

The performance indicators used in this study were the mean absolute error (MAE) and normalized discounted cumulative gain (nDCG), as shown in Eqs.~\ref{eq:mae} and~\ref{eq:ndcg}.
\begin{equation}
    MAE = \frac{1}{|U|} \sum_{i\in U} \frac{1}{|H'_i|} \sum_{j\in H'_i} \left| y_{ij} - \hat{y}_{ij} \right|, \label{eq:mae}
\end{equation}
\begin{equation}
    nDCG_k = \frac{1}{|U|} \sum_{i\in U} \frac{DCG_k^i}{IDCG_k^i}, \label{eq:ndcg}
\end{equation}
\begin{equation}
    DCG_k^i = \sum_{r=1}^k \frac{2^{R_{ir}}-1}{\log_2(1+r)} \nonumber
\end{equation}
where $R_{ir}$, $U$, and $IDCG_k^i$ are the real rating of user $i$ for the $r$-th ranked item, a set of users in the test data and the best possible $DCG_k^i$ for user $i$, respectively.

\subsection{Experimental Results}
We conducted experiments on a non-cold-start scenario and three cold-start scenarios: (1) recommendation of existing items for new users, (2) recommendation of new items for existing users, and (3) recommendation of new items for new users. The first type of cold-start recommendation scenario can be considered as recommendations for new users of a system. The second type can be considered as making a recommendation on new movies or books. The third is a mixture of the first and second cases and is the most difficult recommendation scenario.

\begin{table*}[ht]
\centering
\caption{Experimental results on the MovieLens and Bookcrossing datasets. The value of each cell represents the average value, regardless of the length of the user's item-consumption history.}
\label{tab:exp_results_w_comp}
\begin{tabular}{p{4.5cm} | l | ccc | ccc}
    \hline
    \multirow{2}{4.5cm}{Type}   &\multirow{2}{*}{Method}& \multicolumn{3}{c|}{MovieLens} & \multicolumn{3}{c}{Bookcrossing}        \\ 
    \cline{3-8}
                                &                       & $MAE$   & $nDCG_1$  & $nDCG_3$  & $MAE$   & $nDCG_1$  & $nDCG_3$\\
    \hline \hline
    \multirow{4}{4.5cm}{Recommendation of existing items for existing users}
                                & PPR                   & \textbf{0.1820}   & \textbf{0.9796}   & \textbf{0.9831}  & 3.8092    & 0.8242    & 0.8494 \\ \cline{2-8}
                                & Wide \& Deep          & 0.9047   & 0.9090   & 0.9117  & 1.6206    & 0.9012    & 0.9172 \\ \cline{2-8}
                                & \ours-1                & 0.7661   & 0.8866   & 0.8904  & \textbf{0.7799}    & \textbf{0.9563}    & \textbf{0.9572} \\ \cline{2-8}
                                & \ours-5                & 0.7567   & 0.8870   & 0.8919  & 0.7955    & 0.9546    & 0.9552  \\ 
    \hline \hline
    \multirow{4}{4.5cm}{Recommendation of existing items for new users}
                                & PPR                   & 1.0748   & 0.8299   & 0.8468  & 3.8430    & 0.8201    & 0.8434 \\ \cline{2-8}
                                & Wide \& Deep          & 1.0694   & 0.8559   & 0.8639  & 2.0457    & 0.8238    & 0.8515 \\ \cline{2-8}
                                & \ours-1                & 0.7884   & 0.8799   & 0.8810  & \textbf{1.8701}    & \textbf{0.8265}    & 0.8527 \\ \cline{2-8}
                                & \ours-5                & \textbf{0.7854}   & \textbf{0.8803}   & \textbf{0.8812}  & 1.8767    & 0.8263    & \textbf{0.8532}  \\ 
    \hline \hline
    \multirow{4}{4.5cm}{Recommendation of new items for existing users}
                                & PPR                   & 1.2441   & 0.7289   & 0.7632  &  3.6821    & 0.8115    & 0.8367 \\ \cline{2-8}
                                & Wide \& Deep          & 1.2655   & 0.7420   & 0.7721  & 2.2648    & 0.8190    & 0.8437 \\ \cline{2-8}
                                & \ours-1                & 0.9361   & \textbf{0.7715}   & 0.7990  & \textbf{2.1047}    & \textbf{0.8202}    & \textbf{0.8441} \\ \cline{2-8}
                                & \ours-5                & \textbf{0.9275}   & 0.7697   & \textbf{0.8005}  & 2.1236    & 0.8190    & 0.8440 \\
    \hline \hline
    \multirow{4}{4.5cm}{Recommendation of new items for new users}
                                & PPR                   & 1.2596   & 0.7292   & 0.7634  & 3.7046    & 0.8171    & 0.8381 \\ \cline{2-8}
                                & Wide \& Deep          & 1.3114   & 0.7680   & 0.7874  & 2.3088    & 0.8160    & 0.8405 \\ \cline{2-8}
                                & \ours-1                & 0.9299   & \textbf{0.7760}   & \textbf{0.8011}  & \textbf{2.1475}    & \textbf{0.8184}    & 0.8410 \\ \cline{2-8}
                                & \ours-5                & \textbf{0.9235}   & 0.7752   & 0.8008  & 2.1721    & \textbf{0.8184}    & \textbf{0.8422} \\ 
    \hline
\end{tabular}
\end{table*}

Table~\ref{tab:exp_results_w_comp} shows the performance of the three methods for the four types of recommendation on the two datasets. The recommendation of existing items for existing users represents the performance on the training dataset. The other three types, which represent cold-start scenarios, indicate the test performance in different cases. We denoted the proposed model with $l$ local updates as \ours-$l$. Our model outperforms two comparative methods in three types of cold-start scenarios in the MovieLens and Bookcrossing datasets.
PPR showed the best $MAE$, $nDCG_1$, and $nDCG_3$ in the recommendation of existing items for existing users in the MovieLens dataset, but it showed poor performance in the three cold-start scenarios. From the performance degradation, we could infer that the performance gaps of PPR between the non-cold-start scenario and cold-start scenarios come from the overfitting when sparsity is low. Because the Bookcrossing dataset has large sparsity, the models were not well fitted compared to the MovieLens dataset and it is considered that overfitting did not occur. Moreover, the proposed method performed well when minimal information about users is available (only age), as for the Bookcrossing dataset.
Generally, our model showed good performance in cold-start scenarios, regardless of the amount of user and item information available to the recommender system.

\begin{figure}
  \centering
  \includegraphics[width=78mm]{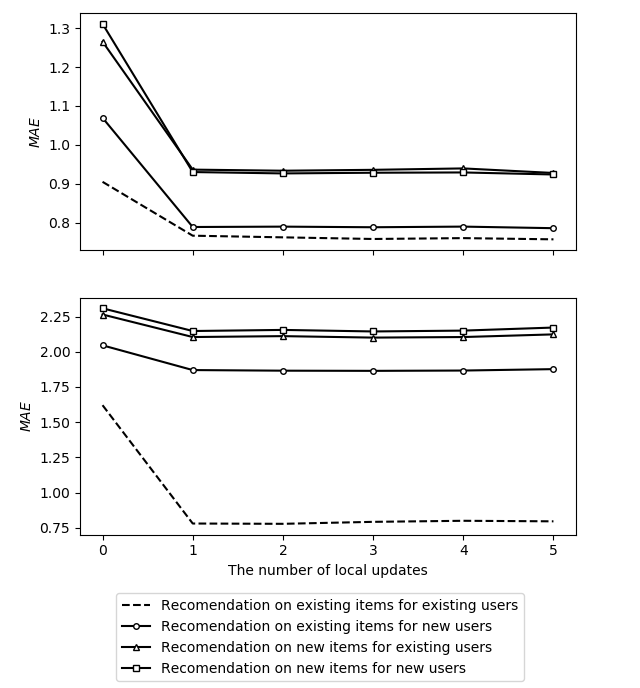}
  \caption{The $MAE$ of our method according to the change in the number of local updates on the (top) MovieLens and (bottom) Bookcrossing datasets.}
  \label{fig:iter_for_per}
\end{figure}

\ours\ achieved remarkable improvements on all types of datasets, even in cases with very few local updates. Figure~\ref{fig:iter_for_per} shows the performance of our method for varying the number of iterations for personalization on the two datasets. The $MAE$ of the proposed method dramatically decreased after the single iteration for all datasets. After one iteration, slight differences were observed for increasing the number of local updates, in contrast to the results from the existing MAML~\cite{finn2017model}, in which the performance improved as the number of iterations increased. 
Our model can adapt quickly to users because a single local update is sufficient. The rapid adaptation allows the proposed method to be applied to online recommendation based on user ratings.

\begin{figure}[t]
  \centering
  \includegraphics[width=80mm]{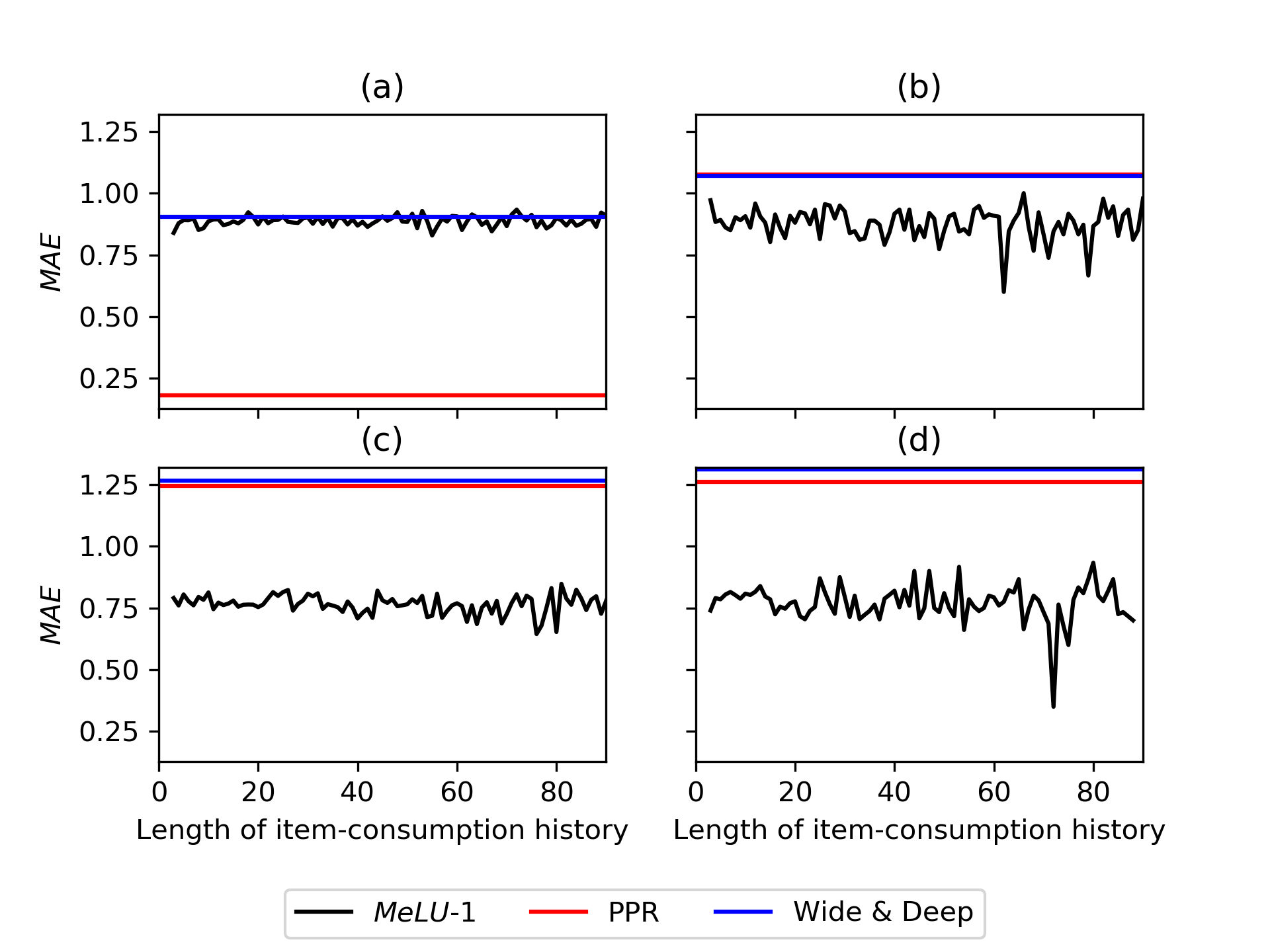}
  \caption{The $MAE$ of our method according to the length of item-consumption history on the MovieLens dataset. (a) Recommendation of existing items for existing users. (b) Recommendation of existing items for new users. (c) Recommendation of new items for existing users. (d) Recommendation of new items for new users.}
  \label{fig:ml_ndcg1}
\end{figure}

\begin{figure}[t]
  \centering
  \includegraphics[width=80mm]{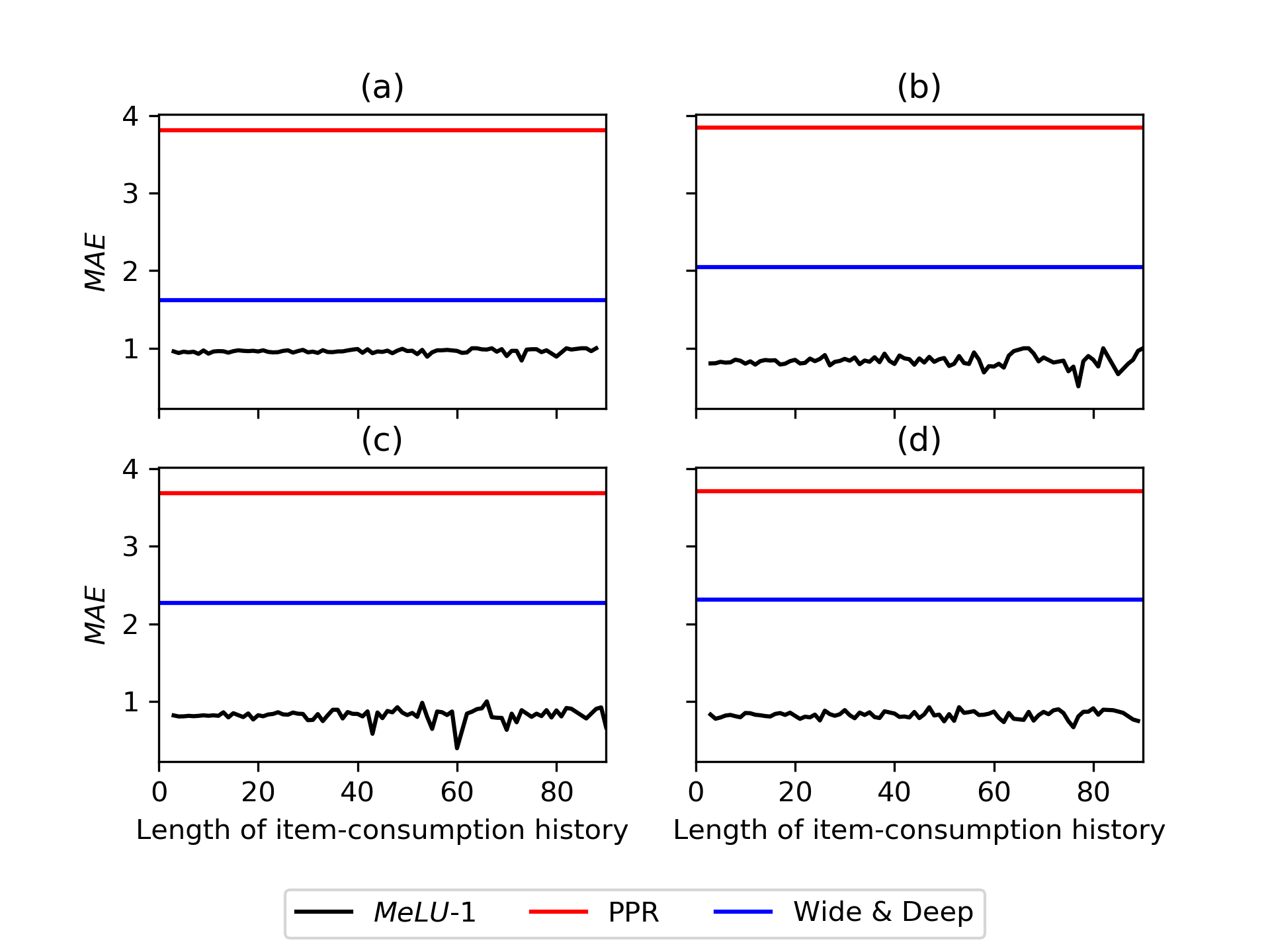}
  \caption{The $MAE$ of our method according to the length of item-consumption history on the Bookcrossing dataset. (a) Recommendation of existing items for existing users. (b) Recommendation of existing items for new users. (c) Recommendation of new items for existing users. (d) Recommendation of new items for new users.}
  \label{fig:bx_ndcg1}
\end{figure}

Figures~\ref{fig:ml_ndcg1} and~\ref{fig:bx_ndcg1} show the $MAE$ of our method according to the length of the item-consumption history when the number of local updates is one. As shown in the figures, the recommendation performances of \ours\ were superb even when the length of the support set was short. Our model showed robust performance with regard to the length of the item-consumption history.
Note that for both datasets, the longer the item-consumption history was, the smaller the number of people. In extreme cases, the number of people was less than five. The performance is believed to be unstable because of insufficient sample size.

\begin{figure*}
  \centering 
  \includegraphics[width=0.875\textwidth]{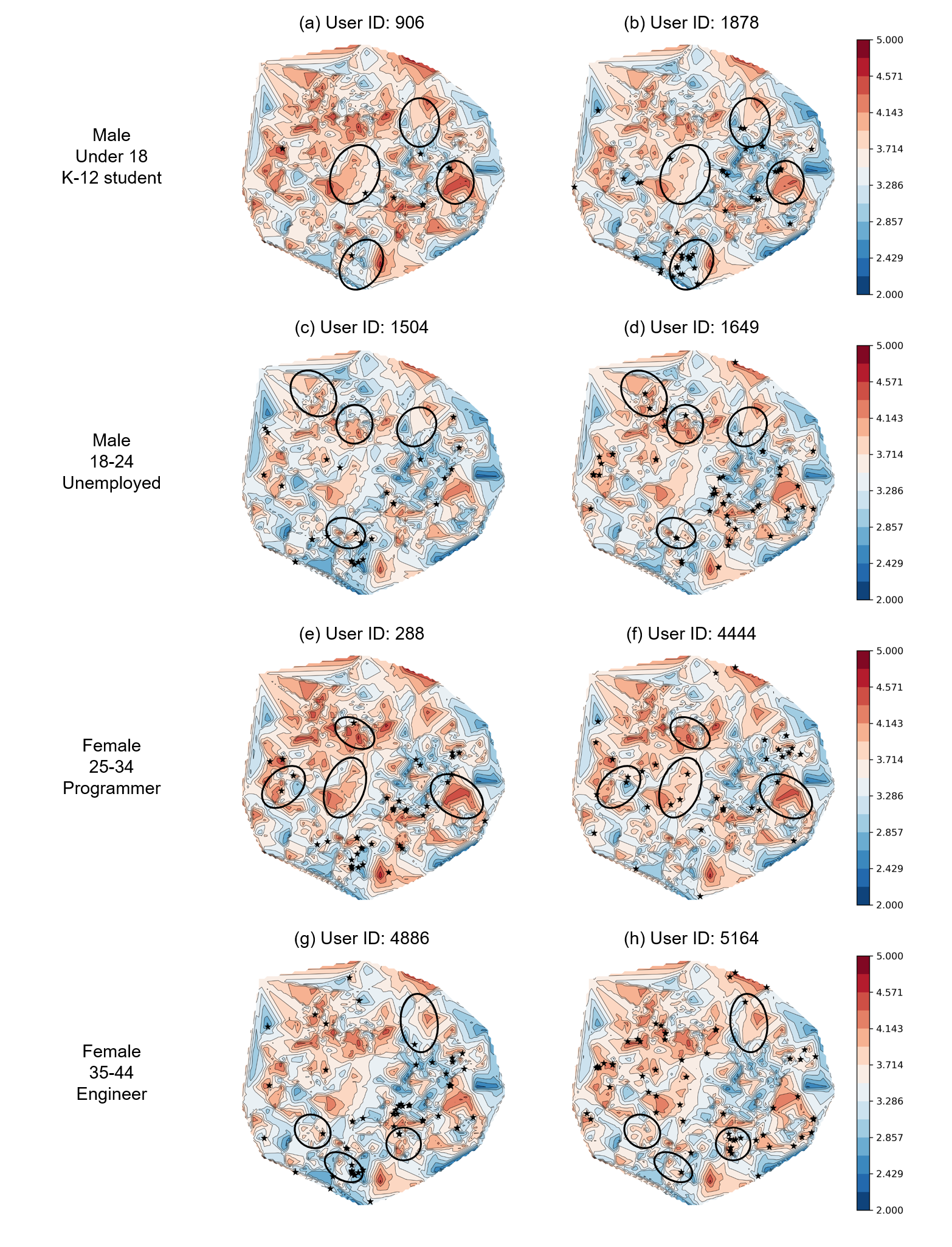}
  \caption{The estimated preference for whole movies by \ours-1 with respect to users. Users like movies located in the red area and dislike movies in the blue area. The black stars represent the movies rated by the user. In each row, the areas with a difference on the contour map are indicated by black circles.}
  \label{fig:casestudy}
\end{figure*}

We qualitatively analyzed the effect of the local update, which aims to provide personalized recommendations. Figure~\ref{fig:casestudy} shows the contour maps of the estimated preferences after one local update for all items of eight MovieLens users. They were divided into four groups, with each group having the same user profiles in each row. For visualization, we used t-SNE~\cite{maaten2008visualizing} to reduce the dimension of the item embeddings into two, which was shared by all figures.
Differences in the ridge and valley lines were observed in the contour maps within groups as well as between groups. The difference between groups could be caused by differences in user profiles. 
The difference within groups verified that the proposed algorithm estimates preferences based on movies belonging to the item-consumption history of each user, even when the user profiles are the same. To summarize, \ours\ estimated the preferences of users within groups as well as users between groups through the local update.

\section{User Study on Evidence Candidates}
\label{sec:exp2}

\subsection{Design of User Study}
\begin{figure}
  \centering
  \includegraphics[width=84mm]{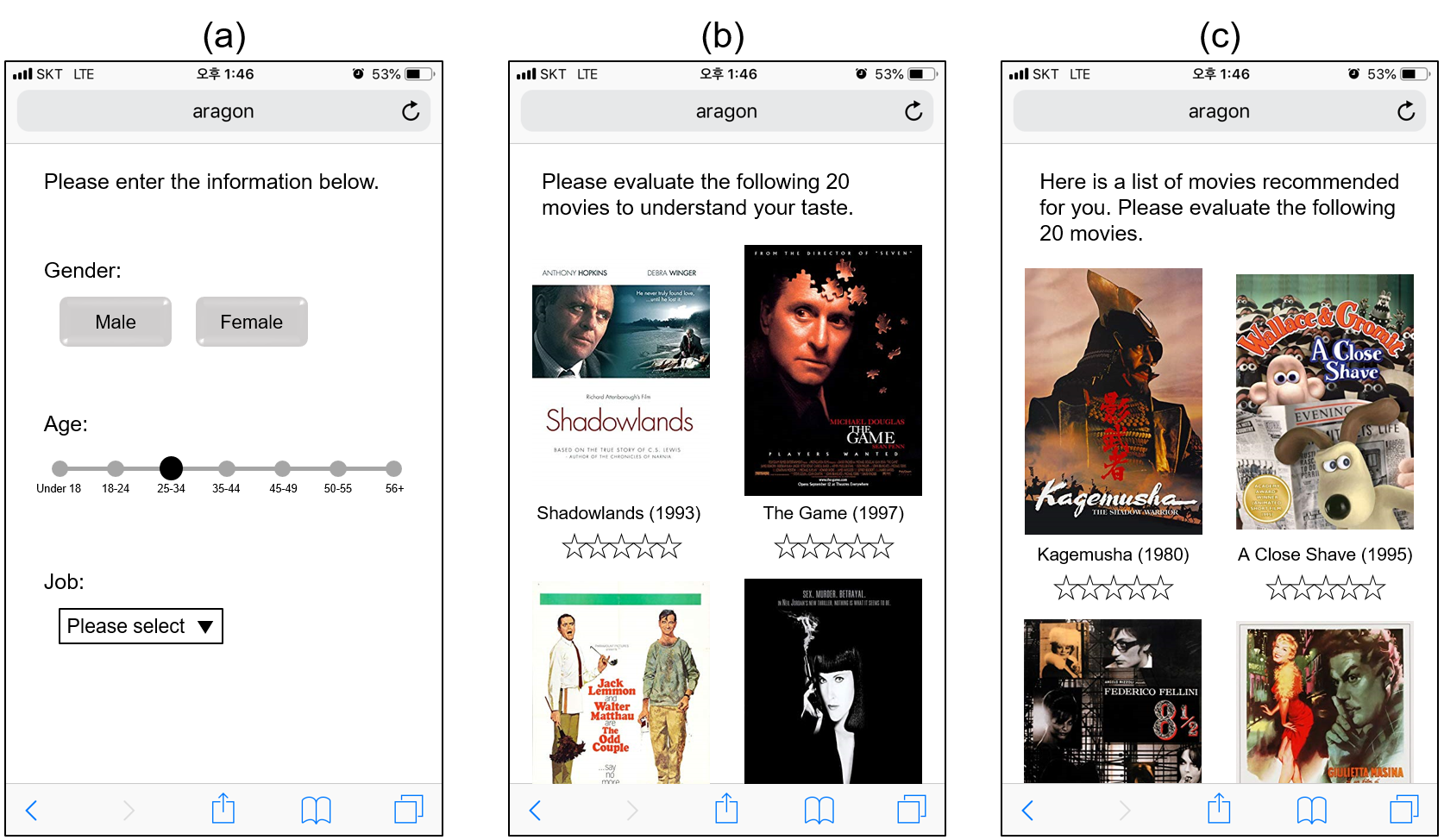}
  \caption{Screenshots of the user survey. (a) First page: survey on user information. (b) Second page: survey on evidence candidates. (c) Third page: survey on recommended items.}
  \label{fig:user_study_demo}
\end{figure}

We conducted user study experiments to compare the proposed and popularity-based evidence candidate selection strategy.
In this section, we considered only the MovieLens dataset because more people frequently watch movies than read books. We created the demonstration page for the user study shown in Figure~\ref{fig:user_study_demo}. On the first page, as in MovieLens, we collected the user's basic information, such as gender, age, and occupation. On the second page, we presented twenty movies to the user and then obtained item preferences on a five-point Likert scale, which means that we selected twenty evidence candidates. At this time, the movies the user is not familiar with were marked as `I do not know this movie.' These movies were not included in the user's item-consumption history. We randomly showed one of the popularity-based or our strategy-based movie lists as an A/B test; we listed the two movie lists in Appendix~\ref{sec:appendix}. 
When calculating the scores for both evidence candidates, we used existing user-item pairs, and calculated gradients from the model with one local update (\ie, \ours-1). On the third page, accepting the movies rated by the user among the twenty movies, our model presented new twenty movies that the user might like. We only considered the one-round interaction to verify the effectiveness of extreme user cold-state, but it could be expanded into a multiple-round interactions with the simple expansion on the third page.
Additionally, we collected the ratings of the recommended items on a five-point Likert scale, as on the second page. Based on the survey, we evaluated which movie list more accurately discriminated user preferences. 

\subsection{Results of User Study}
We quantitatively validated the evidence candidate selection via a user study. We surveyed eighty-four users in total. Forty-five of the users rated movies extracted by popularity only, and the others rated movies extracted using our strategy. Table~\ref{tab:suvery} shows the summary of the survey results. Naturally, on the second page, our evidence candidates were rarely rated by users compared to the popularity-based evidence candidates. While approximately one-half of the popularity-based evidence candidates were chosen, only one-quarter of our evidence candidates were selected. However, on the third page, the proposed strategy showed better performance on all indicators. The respondents chose more movies, and the average ratings and $nDCG_1$ of the suggested movies were higher. Note that the proposed strategy used smaller evidences than popularity-based strategy for the local update, although the average ratings on the evidences were similar. Therefore, our strategy provides reliable evidence candidates and can quickly identify individual preferences of new users for the items.

We interviewed some of the participants to qualitatively analyze the users' response to the system. The reactions from two groups conflicted. One user assigned to policy A (\ie, popularity-based evidence) said that she could made a lot of choices among the movies on the second page because she knew most of them, but she could not chose any recommended movies on the third page. Another user assigned to the same policy answered that the recommended movies did not seem to grasp his taste. By contrast, a user assigned to policy B (\ie, proposed strategy-based evidences) responded that it was nice to have more preferred movies on the third page than on the second page.

\begin{table}[]
\centering
\caption{Survey results on evidence candidate selection. On the left side of the table, we denote the evidence candidates as EC, and the recommended items as RI, which were used to calculate the measures of the corresponding rows.}
\label{tab:suvery}
\begin{tabular}{ll|c|c}
    \cline{2-4}
    & \textbf{Strategy}           & \textbf{Popularity-based} & \textbf{Ours} \\ 
    \cline{2-4}
    & Number of users     & 45            & 39\\ 
    \cline{2-4}
    \multirow{2}{*}{\rot{EC}} & Avg. number of selection     & 10.89             & 5.31 \\ 
    & Avg. ratings     & 4.02             & 3.96 \\ 
    \cline{2-4}
    \multirow{3}{*}{\rot{RI}} & Avg. number of selection     & 3.29             & 4.28 \\ 
    & Avg. ratings     & 3.94             & 4.29 \\ 
    & Avg. $nDCG_1$     & 0.2756             & 0.3692 \\ 
    \cline{2-4}
\end{tabular}
\end{table}

\section{Conclusion}
\label{sec:con}

In this paper, we proposed a MAML-based recommender system that can identify personalized preferences. Our model was able to estimate user preferences based on only a small number of items. We found that the proposed method outperforms two methods on two benchmark datasets. Moreover, we qualitatively found that the local update using item-consumption history effectively identifies user preferences. In addition, based on the local update, we devised a strategy to select evidence candidates. As a result of an A/B test with eighty-four users, the candidates of the proposed strategy were selected less often than those of the popularity-based strategy, but the users were more satisfied with the recommendation results by the proposed strategy-based evidences.

Three promising topics for future research remain. First, the model update cycle must be verified before real-world applications. Although meta-learning is good at learning new things, the performance may not be guaranteed when an entirely new type of user appears in the recommender system.
Second, future studies should explore the variants of evidence candidate selection strategy. 
When calculating the scores for evidence candidate selection, the weighted summation or the addition of other factors may yield better results.
Third, the variants of meta-learning based recommender system should be studied. For example, meta-learning based collaborative filter could be designed.

\bibliographystyle{ACM-Reference-Format}

\onecolumn
\appendix
\counterwithin{table}{section}
\section*{APPENDIX}
\setcounter{section}{0}
\section{Lists of evidence candidates.}
\label{sec:appendix}

\begin{table}[h]
\centering
\caption{Comparison of popularity-based and proposed strategy-based evidence candidates. All movies are sorted by the score of each method. Note that only three movies are overlapped (marked in bold).}

\label{tab:appendix}
\begin{tabular}{c||c|c||c|c}
\hline
     \multirow{2}{*}{Rank}& \multicolumn{2}{c||}{\textbf{Popularity-based strategy}}                                                                   & \multicolumn{2}{c}{\textbf{\begin{tabular}[c]{@{}c@{}}Our strategy\end{tabular}}}                  \\ \cline{2-5}
 & Title                                                                                                  & Score   & Title                                                                                                  & Score    \\ \hline\hline 
1    & \begin{tabular}[c]{@{}c@{}}Star Wars: Episode IV - A New Hope (1977)\end{tabular}                  & 1.00000 & Shadowlands (1993)                                                                                     & 0.05576 \\ \hline
2    & \small \begin{tabular}[c]{@{}c@{}}Star Wars: Episode V - The Empire Strikes Back (1980)\end{tabular}      & 0.93272 & The Game (1997)                                                                                        & 0.01032 \\ \hline
3    & \small \textbf{\begin{tabular}[c]{@{}c@{}}Star Wars: Episode VI - Return of the Jedi (1983)\end{tabular}} & 0.89677 & The Odd Couple (1968)                                                                                  & 0.01021 \\ \hline
4    & \textbf{Terminator 2 (1991)}                                                                           & 0.83963 & The Crying Game (1992)                                                                                 & 0.00846 \\ \hline
5    & Jurassic Park (1993)                                                                                   & 0.80415 & 2010 (1984)                                                                                            & 0.00375 \\ \hline
6    & The Silence of the Lambs (1991)                                                                        & 0.78295 & \textbf{L.A. Confidential (1997)}                                                                      & 0.00282 \\ \hline
7    & Braveheart (1995)                                                                                      & 0.76866 & Contact (1997)                                                                                         & 0.00224 \\ \hline
8    & Fargo (1996)                                                                                           & 0.76544 & Copycat (1995)                                                                                         & 0.00216 \\ \hline
9    & Back to the Future (1985)                                                                              & 0.74332 & Jaws (1975)                                                                                            & 0.00128 \\ \hline
10   & Raiders of the Lost Ark (1981)                                                                         & 0.73871 & The Terminator (1984)                                                                                  & 0.00121 \\ \hline
11   & Schindler's List (1993)                                                                                & 0.73733 & Star Trek: First Contact (1996)                                                                        & 0.00119 \\ \hline
12   & Men in Black (1997)                                                                                    & 0.72581 & \small \textbf{\begin{tabular}[c]{@{}c@{}}Star Wars: Episode VI - Return of the Jedi (1983)\end{tabular}} & 0.00118 \\ \hline
13   & \textbf{L.A. Confidential (1997)}                                                                      & 0.72442 & Beauty and the Beast (1991)                                                                            & 0.00102 \\ \hline
14   & The Godfather (1972)                                                                                   & 0.67742 & \textbf{Terminator 2 (1991)}                                                                           & 0.00090 \\ \hline
15   & The Shawshank Redemption (1994)                                                                        & 0.67051 & Manhattan (1979)                                                                                       & 0.00083 \\ \hline
16   & The Princess Bride (1987)                                                                              & 0.66728 & Twelve Monkeys (1995)                                                                                  & 0.00073 \\ \hline
17   & Groundhog Day (1993)                                                                                   & 0.61429 & Clerks (1994)                                                                                          & 0.00070 \\ \hline
18   & Forrest Gump (1994)                                                                                    & 0.61014 & Big (1988)                                                                                             & 0.00064 \\ \hline
19   & Pulp Fiction (1994)                                                                                    & 0.59724 & Emma (1996)                                                                                            & 0.00062 \\ \hline
20   & E.T. the Extra-Terrestrial (1982)                                                                      & 0.59078 & Die Hard (1988)                                                                                        & 0.00057 \\ \hline
\end{tabular}
\end{table}

\end{document}